\documentclass[%
 reprint,
superscriptaddress,
 amsmath,amssymb,
 aps,
 pre
]{revtex4-2}

\usepackage[dvipsnames]{xcolor}
\usepackage{graphicx}
\usepackage{dcolumn}
\usepackage{bm}
\usepackage{cleveref}

\begin{document}

\preprint{APS/123-QED}

\title{Ordered buckling structures in a twisted crimped tube}

\author{Pan Dong}
\affiliation{Department of Physics, Syracuse University, Syracuse, NY 13244}
\affiliation{BioInspired Institute, Syracuse University, Syracuse, NY 13244}
\author{Nathan C. Keim}
\email{keim@psu.edu}
\affiliation{Department of Physics, The Pennsylvania State University, University Park, PA 16802}
\author{Joseph D. Paulsen}
\email{paulse8@stolaf.edu}
\affiliation{Department of Physics, Syracuse University, Syracuse, NY 13244}
\affiliation{BioInspired Institute, Syracuse University, Syracuse, NY 13244}
\affiliation{Department of Physics, St.~Olaf College, Northfield, MN 55057}
\date{\today}

\begin{abstract}
When a ribbon or tube is twisted far enough it forms buckles and wrinkles. Its new geometry can be strikingly ordered, or hopelessly disordered. Here we study this process in a tube with hybrid boundary conditions: one end a cylinder, and the other end crimped flat like a ribbon, so that the sample resembles a toothpaste tube. The resulting irregular structures and mechanical responses can be dramatically different from those of a ribbon. However, when we form two creases in the tube prior to twisting, we obtain an ordered structure composed of repeating triangular facets oriented at varying angles, and a more elastic torque response, reminiscent of the creased helicoid structure of a twisted ribbon. We measure how the torque and structural evolution depend on parameters such as material thickness and the twist angle. When only part of the tube is pre-creased, the ordered structures are confined to this segment. Surprisingly, in some tubes made from thicker sheets, an ordered structure forms without pre-creasing. This study provides insights into controlling the buckling of thin shells, offering a potential pathway for designing ordered structures in soft materials.

\end{abstract}

\maketitle

\section{Introduction}

Buckled structures are ubiquitous in sheets and surfaces, from  smooth wrinkles in taut bed sheets to the sharp creases and points on a crushed soda can. 
These patterns form readily in response to external forces or torques, particularly in thin sheets which bend much more easily than they stretch \cite{Milner1989, Cerda2003, Seffen2013, Bella2017}. 
Even when a sheet is manipulated only at its edges, the resulting range of buckling patterns prompts fundamental questions: What geometries are possible, and in what ways do boundary conditions control their formation?

To understand how the boundary conditions control the buckling of thin sheets, prior studies have examined twisted ribbons, a simple yet informative model system~\cite{Korte2010,Bohr2013,Chopin2013,Chopin2014,Chopin2016,Dinh2016,Fosdick2016,Kohn2018,Demery2018,Chopin2022,Leembruggen2024}.
The combination of axial twist and tension allow one to navigate a number of striking post-buckled morphologies. 
Thin tubes---representing another fundamental geometric configuration---have been shown to exhibit diverse buckling patterns when subjected to different boundary conditions~\cite{Hamm2004,Maha2007,Santangelo2013,Seffen2014,Pan2023,Lu2025}.
For instance, a twisted cylindrical sheet can show a distinct pattern of wrinkles \cite{Pan2023}, with a symmetry that matches the circular shape of its boundaries. These observations suggest that the cross-sections of the boundaries might play an important role in controlling buckling patterns in twisted sheets. 
For ribbons and tubes, the boundaries are defined by linear and circular cross-sections. 
What arises when a thin sheet is constrained by mixed boundary conditions—specifically, one linear edge and one circular edge? Can this geometry give rise to new structures under twist, or will its behavior mirror that observed in ribbons or tubes?

\begin{figure*}[ht]
\centering
\includegraphics[width=\textwidth]{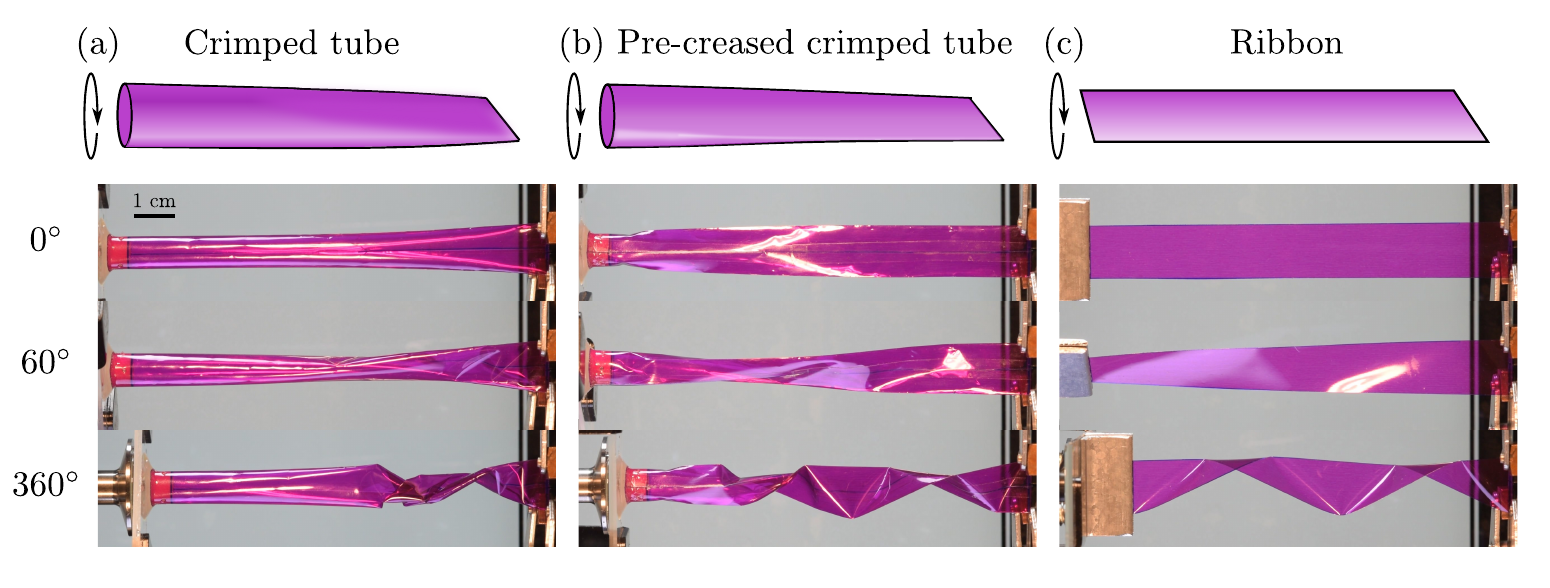}
\caption{\label{fig:360_twist} \textbf{Buckling morphologies of three slender structures under tension and twist.} 
All samples have $t=38$~$\mu$m, $W=13.5$~mm, $L=108$~mm. 
(a) A crimped tube evolves into a disordered twisted structure, under constant tension $T=0.5$ N. 
(b) A pre-creased crimped tube develops an ordered creased helicoid structure, emanating from the crimped end, for $T=0.5$ N. 
The structure is reminiscent of a buckling pattern in ribbons \cite{Korte2010}. 
(c) Twisted ribbon, for comparison with the crimped tube. We set $T=0.25$ N for the ribbon as the emergent structure in panel (b) is effectively a double-layered ribbon.  Figure~\ref{fig:sm1} shows the same sequences with increments of $30^\circ$.
}
\end{figure*}

To explore these questions, here we crimp one circular end of a cylindrical sheet into a flat ribbon, so that the sample resembles a toothpaste tube, and we observe its morphology and mechanics under controlled tension and axial twist. 
We perform torque measurements during an axial twist with a fixed vertical tension. 
Our samples generally undergo a seemingly random crumpling process [Fig.~\ref{fig:360_twist}(a)], consistent with a twisted cylindrical tube \cite{Dawadi2024}. 
But, if we apply two creases along the edge of the tube, the sheet instead forms a simple repeating pattern of triangular facets [Fig.~\ref{fig:360_twist}(b)]. 
This response resembles not only the morphology of a twisted ribbon [Fig.~\ref{fig:360_twist}(c)], but also its mechanics  (Fig.~\ref{fig:torque}). 
We investigate how the tension, thickness, twist angle, and length of the pre-creases affect these results. 
Finally, we discover a set of parameters where ordered buckling can arise without pre-creases. 
Our results identify routes to controlling the buckling patterns of tubes manipulated at their ends, which could be useful for the design and control of mechanical metamaterials built from thin-walled tubes.

\section{Experimental Setup}

\textit{Fabrication.---} 
We construct our samples from thin color-coded polyester shim stock (Artus Corporation) with thickness $t$ ranging from $38$ to $127$~$\mu$m. 
We measured the Young's modulus, $E$, of the 38~$\mu$m-thick sheet by clamping a $13$~mm-wide, 51~mm-long strip in a rheometer (Anton Paar MCR302-WESP) and measuring the slope of its stress versus strain for small extensions. 
We find $E=2.6$~GPa, which is within the expected range for polyester. 

To form each sample, we cut an initially flat sheet into a rectangle, roll it into a cylindrical tube, and crimp it at one end, forming a familiar ``toothpaste tube" shape.
We use double-sided tape and epoxy to attach the cylindrical end to the narrow section of a sleeve washer (Essentra Components)---an 
$8.4$~mm diameter cylinder extending from a larger plastic annulus. 
We seal the crimped end with double-sided tape, and we seal the seam along the length of the tube using 5~mm-wide clear tape. 
The seam meets the crimped end midway between the two points where the sheet is pinched by the clamp. 
To ensure that the enclosed volume is free to vary during the experiment, we drill holes through the sample at the location of the sleeve washer to let air flow in and out of the tube. 

We use a standardized sample width of $W=13.5$~mm,  measured at the crimped end of the tube and anywhere along the ribbon's length. 
The length of the unsupported portion of the samples is $L=108$~mm, except where otherwise stated. 

\textit{Mechanical testing.---} 
We seek to apply controlled tension to the sample while varying the twist angle, following the mechanical protocols explored recently for a ribbon \cite{Korte2010,Chopin2014}. 
Using the rheometer, we are able to additionally measure the time-dependent torque exerted on the sample throughout the experiment. 

To mount our sample to the rheometer, we attach the sleeve washer to a glass microscope slide (Fisher Scientific, $25\times75\times1$~mm) with epoxy. We affix a second glass slide to the parallel-plate rheometer tool using hot-melt adhesive, and we use two mini C-clamps to secure the two glass slides together. 
This allows for precise alignment of the axis of rotation of the rheometer with the axis of the sleeve washer. 
We clamp the crimped end of the tube between two ``L"-shaped base clamps. 
By removing the base of the rheometer, we are able to mount these clamps on an optical post that we affix directly to the optical table supporting the rheometer. 
The whole setup is shown in Fig.~\ref{fig:setup}. 

Once mounted, we perform angle-controlled rotation under a small, fixed tension. 
The rotational velocity is fixed to $0.1$~degrees/s, and we record movies of the experiments with a DSLR camera (Nikon D5300). 
The samples are viewed directly from the side, with the camera approximately 1 m from the sample to minimize distortion. 
We note that the nominal material strain is very small in our experiments, $T/(E t W) \simeq 3.7 \times 10^{-4}$. 
Nevertheless, the deformations may be large, due to large rotations within these slender structures.

\section{Results}

\textit{Buckling morphology.---} We now explore how the geometric structure of a pristine crimped tube evolves when it is twisted to $360^{\circ}$ under constant tension $T=0.5$~N. 
Figure~\ref{fig:360_twist}(a) shows snapshots at $60^\circ$ and $360^{\circ}$. 
The portion of the sample near the crimped end becomes crumpled; it develops sharp deformations joined by ridges in varying orientations. 
This response is reminiscent of a thin twisted cylinder~\cite{Dawadi2024}. 
The remaining portion of the sample near the circular end remains relatively intact with no creases. 

Looking for other possible responses, we repeated the experiment with a fresh sample, but this time we applied two creases on the edge of the un-twisted tube by sliding a hair clip along its length. 
Surprisingly, a pre-creased crimped tube subjected to the same twist formed a relatively ordered structure [Fig.~\ref{fig:360_twist}(b)]: 
we observe many flat triangular facets, emanating from the crimped side of the tube. 

These creases and facets are reminiscent of the ``creased helicoid'' buckling pattern that can form in a twisted ribbon \cite{Korte2010,Chopin2013,Bohr2013}. 
We reproduce this response in Fig.~\ref{fig:360_twist}(c) with a ribbon cut from the same plastic stock. 
To clamp the ribbon, in place of the sleeve washer we use two pieces of L-shaped stainless steel attached to a glass slide with epoxy. Then we fix the ribbon to the steel with two C-clamps.

\begin{figure}[b]
\includegraphics[width=0.9\columnwidth]{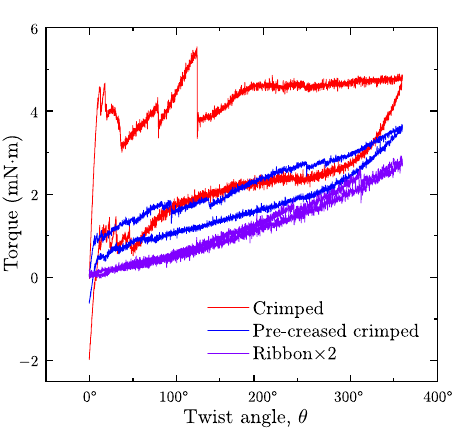}
\caption{\label{fig:torque}
\textbf{Mechanical response of the three
samples in Fig.~\ref{fig:360_twist}.} 
Samples are twisted from $0^\circ$ to $360^\circ$ and back to $0^\circ$. 
Because the ribbon width is half the circumference of the tubes, we plot twice the value of the measured torque for the ribbon. 
The data show that pre-creasing a crimped tube brings its mechanical response close to that of a ribbon. 
}
\end{figure}

\textit{Mechanical response.---} 
Comparing the three cases in Fig.~\ref{fig:360_twist} prompts us to ask whether pre-creasing the tube merely gives it a more ribbon-like appearance, or if it signals a qualitative change in the physics of deformation. 
To test this possibility, in Fig.~\ref{fig:torque} we measure the torque of each sample during twisting and untwisting. 
The data for the uncreased tube show intermittent, large drops in the torque, consistent with the formation of crumpled structures, such as d-cones and minimal ridges, and snap-through events involving these structures \cite{Croll2019,Shohat2022,Shohat2023}. 
By contrast, the data for the pre-creased tube show much smaller drops, reduced hysteresis, and markedly lower torques. 
Although the exact location and size of the drops varies from sample to sample, these overall trends are robust. 
We show data from fresh samples twisted to $240^\circ$, $360^\circ$, and $450^\circ$ in the appendix (Fig.~\ref{fig:3twists}). 

As a second point of comparison, in Fig.~\ref{fig:torque} we also show the torque from the ribbon of Fig.~\ref{fig:360_twist}(c), multiplied by two since the ribbon width is half the circumference of the tubes. 
The ribbon data are consistent with recent measurements by Chopin and Filho~\cite{Chopin2019} of a ribbon at low tension twisted far past the buckling threshold; in the appendix (Fig.~\ref{fig:3twists}) we compare our data to those authors' theoretical prediction for ribbons at higher tension. 
Here, data from the ribbon and the pre-creased tube show a qualitative resemblance. 
These findings strengthen the notion that pre-creasing transforms the crimped tube’s mechanical response, making it behave more like a ribbon than a tube. 

\begin{figure}[t]
\centering
\includegraphics[width=1.0\columnwidth]{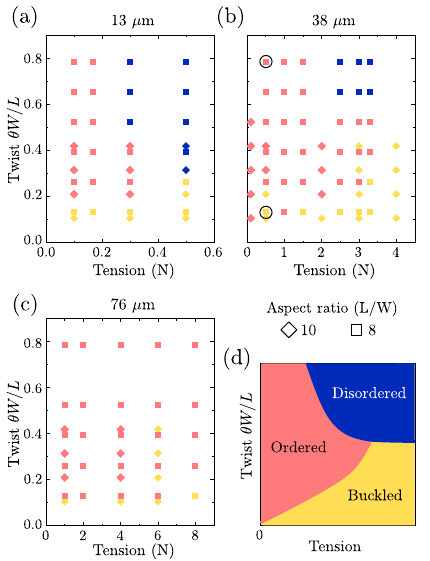}
\caption{\label{fig:diagram} 
\textbf{How the morphology varies with twist, tension, and thickness.} 
(a--c) Each panel shows a different sheet thickness. 
The vertical axis is the twist angle $\theta$, scaled by $W/L$ to convert to the pitch of an equivalent helix. 
Results are robust under a variation of the aspect ratio, indicated by the symbol shape. Open circles in (b) indicate the 60$^\circ$ and 360$^\circ$ images in Fig.~\ref{fig:360_twist}(b).
(d) Qualitative phase diagram that labels the colors used in (a--c), and that captures the results for all three thicknesses. 
}
\end{figure}

\begin{figure}[b]
\centering
\includegraphics[width=0.95\columnwidth]{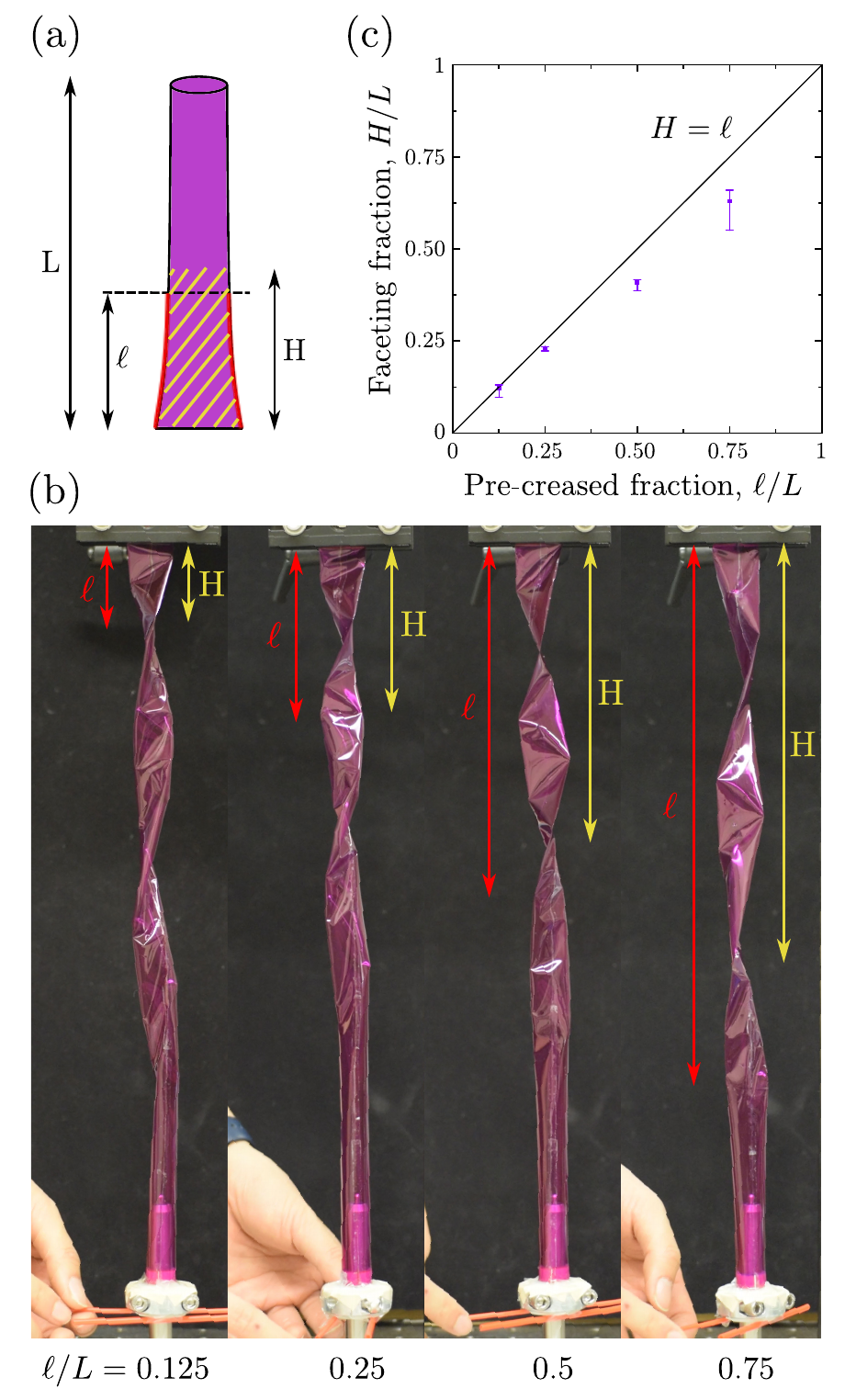}
\caption{\label{fig:partial_creasing} 
\textbf{Buckling of four twisted crimped tubes that are partially pre-creased.}
(a) Schematic of a crimped tube with a fraction $\ell/L$ pre-creased (red lines along its sides), which leads to ordered facets in a fraction $H/L$ (shaded region). 
(b) Photos of samples twisted to $360^\circ$ with pre-creased fractions $\ell/L=0.125$, $0.25$, $0.5$, and $0.75$. 
The samples have $t=38~\mu$m, $W=25.5$~mm, and $L=408$~mm, and the crimped end is at the top. 
(c) We measure the length $H$ of the region that develops ordered facets when each tube is twisted. 
Error bars represent the uncertainty in assessing the end of the ordered faceted region. 
}
\end{figure}

\textit{Generality of ribbon-like response.---} 
As we now show, the similarity between the pre-creased tube and the ribbon is evident under a range of parameters. 
We prepared pre-creased crimped tubes with $t=13$, $38$, and $76~\mu$m, and twisted them from $0^{\circ}$ to $360^{\circ}$ while looking for consecutive flat triangular facets with orientations in a helical pattern,  like the twisted ribbon in Fig.~\ref{fig:360_twist}(c). 
We kept the same width at the crimped end ($W=13.5$~mm) and used $L=108$~mm and $L=135$~mm to achieve two aspect ratios, $L/W = 10$ and 8. 
Figure~\ref{fig:diagram}(a--c) plots the morphologies seen for these thicknesses, at a range of tensions and twisting values $\theta W/L$ (a proxy for the slope of the creased edge). 
For each aspect ratio, each vertical column of points is from a single fresh sample, where we monitor the evolving morphology as we vary $\theta$. 
We identify three distinct morphologies, indicated by the colors that are labeled in Fig.~\ref{fig:diagram}(d): regular triangular facets (``Ordered''); small numbers of shallow creases (``Buckled'') as in the middle row of Fig.~\ref{fig:360_twist}(a, b); and ``Disordered'' shapes that are strongly creased, often with self-contact, as in the bottom image of Fig.~\ref{fig:360_twist}(a). For a given thickness and vertical tension, ordered triangular facets arise over a range of twist angles. As the vertical tension increases, there is a narrower range of twist angles over which triangular facets can be observed. 

Based on these observations, we draw a possible  phase diagram of twisted crimped tubes in Fig.~\ref{fig:diagram}(d). 
The ``Ordered'' region has roughly the same shape and placement as the creased helicoid region on the phase diagram of a twisted ribbon~\cite{Chopin2013}, using the same axes, further suggesting that the resemblance in Fig.~\ref{fig:360_twist}(b, c) is more than superficial. Interestingly, when the thickness is $76~\mu$m, we do not find a disordered structure over the range of accessible parameters [Fig.~\ref{fig:diagram}(c)]. We revisit the case of thick-walled tubes below.

\begin{figure*}[ht!]
\centering
\includegraphics[width=1.0\textwidth]{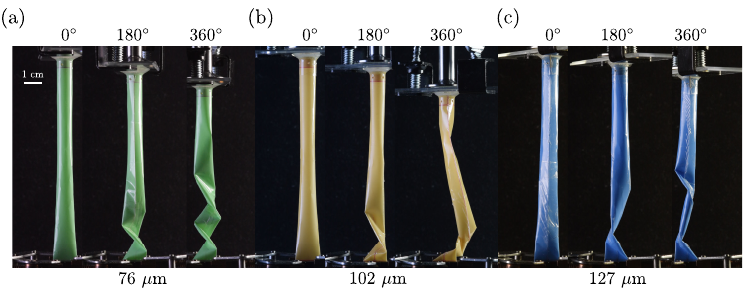}
\caption{\label{fig:uncreased}
\textbf{Buckling morphology of thick, crimped tubes with no pre-creasing.}
All samples have $W=13.5$~mm and $L=108$~mm. 
The thicknesses are: (a) $t=76~\mu$m,
(b) $t=102~\mu$m,
(c) $t=127~\mu$m, and the normal forces are $T = 1$~N, 1.33~N, and 1.67~N, in proportion to thickness. 
The $76~\mu$m tube develops ordered facets, despite the absence of pre-creasing to direct the buckling pathway. 
}
\end{figure*}

\textit{Partial pre-creasing.---} 
Do ordered facets arise only from local pre-creasing of the edges of the tube, or are there long-range effects? To explore this question, in Fig.~\ref{fig:partial_creasing} we twist samples that have been pre-creased for only a fraction of their length, starting from the crimped end. To better resolve local effects, the samples are wider ($W=25.5$~mm) and much longer ($L=408$~mm), so that they do not fit in our rheometer setup and require a separate apparatus. 
We hang a collar clamp at the bottom end of the tube, attaching it to the bottom of the sleeve washer with double-sided tape. 
This collar serves two purposes: (i) the weight of the collar provides a consistent tension of 0.75~N during the twist, and (ii) the collar fits around a post that provides a consistent axis of rotation, as the tube is twisted by hand. 
Because this post blocks air from freely flowing out the end of the tube, here we puncture many small holes along the length of the tube to allow the enclosed volume to change during the deformations. 

We vary the pre-creased fraction from $12.5\%$ to $75\%$ in Fig.~\ref{fig:partial_creasing}(b) and measure the length $H$ of the region where distinct facets form when the tube is twisted to $360^{\circ}$ [Fig.~\ref{fig:partial_creasing}(c)]. 
We observe that facets can only form in the pre-creased region, and that they do not entirely fill the pre-creased region when the pre-creasing boundary approaches the circular end of the tube. 
These results suggest that the buckling pathways in twisted tubes are responsive to local conditions.

\textit{Thicker tubes.---} 
In Figs.~\ref{fig:360_twist} and \ref{fig:partial_creasing}, ordered faceting was only observed where the tube is pre-creased. However, as we now show, thicker tubes may form ordered facets even when they are not pre-creased. 
Figure~\ref{fig:uncreased} shows further experiments on un-creased crimped tubes with thicknesses 76~$\mu$m, 102~$\mu$m, and 127~$\mu$m. 
Remarkably, the 76$~\mu$m tube gradually displays more ordered triangular facets as it is twisted [Fig.~\ref{fig:uncreased}(a)]. The 102~$\mu$m and 127~$\mu$m tubes have similar behavior at 180$^{\circ}$, but self-contact destroys the ordered pattern at 360$^\circ$ [Fig.~\ref{fig:uncreased}(b) and (c)].

\section{Discussion}
We have studied how boundary conditions affect the formation of ordered and disordered structures in twisted thin-walled tubes. 
Our results show that a pre-creased crimped tube can behave more like a ribbon than a tube, in both its geometrical and mechanical response. 
By changing vertical tension, twist angle, and thickness, the twisted crimped tube can transition between buckled, ordered, and disordered states, in a manner that resembles the phase diagram of a twisted ribbon. 
This control appears to be local: Pre-creasing a portion of the tube enables ordered facets within this region only. 
Our findings thus demonstrate multiple avenues for controlling such buckling pathways, and they reveal a rich qualitative landscape of phenomena on a seemingly simple crimped tube. 

Opportunities for more quantitative study include the prominent role of sheet thickness in Fig.~\ref{fig:uncreased}. 
There, the 76~$\mu$m-thick crimped tube shows perfect triangular facets when twisted, even without pre-creasing, reminiscent of self-folded origami~\cite{Santangelo2017}. 
Yet, as thickness increases further, the buckling becomes disordered once again. 
We note that in other contexts, sheet thickness sets the relative ease of buckling the intact sheet in a new place (i.e.\ forming a d-cone) compared to continuously deforming an existing buckled structure~\cite{Timounay2020, Hutton2024}. 
This leads us to speculate that if the tube is too thin, d-cones and new creases can proliferate as their energetic cost is relatively smaller, making ordered facets perhaps less likely. 
If the tube is too thick, the smaller number of d-cones would thus tend to be farther apart; if the average spacing between d-cones is comparable to the tube length [Fig.~\ref{fig:uncreased}(b, c)], this would be incompatible with ordered facets. 
As a tube with intermediate thickness is twisted, as in Fig.~\ref{fig:uncreased}(a), the formation of d-cones at the curved sides of the tube seems to match the development of ordered facets, mimicking  the creased helicoid structure of a twisted ribbon. 
While many of our results can be connected with the behavior of ribbons, the rich behaviors of the uncreased case represent a much more challenging combination of kinetics, geometrically-nonlinear mechanics, and boundary conditions.

\begin{acknowledgments}
We thank David Pratt and Andrew Nadlman in the Syracuse University Physics Machine Shop for assistance with the sample fabrication and clamp design. 
Funding support from NSF-DMR-2318680 (P.D.~and J.D.P.) is gratefully acknowledged. 
\end{acknowledgments}

\section*{Data availability}
The data that support the findings of this article are openly available \cite{Dong2025_Dataset}.

\appendix
\section{}
\begin{figure*}[]
\includegraphics[width=0.85\textwidth]{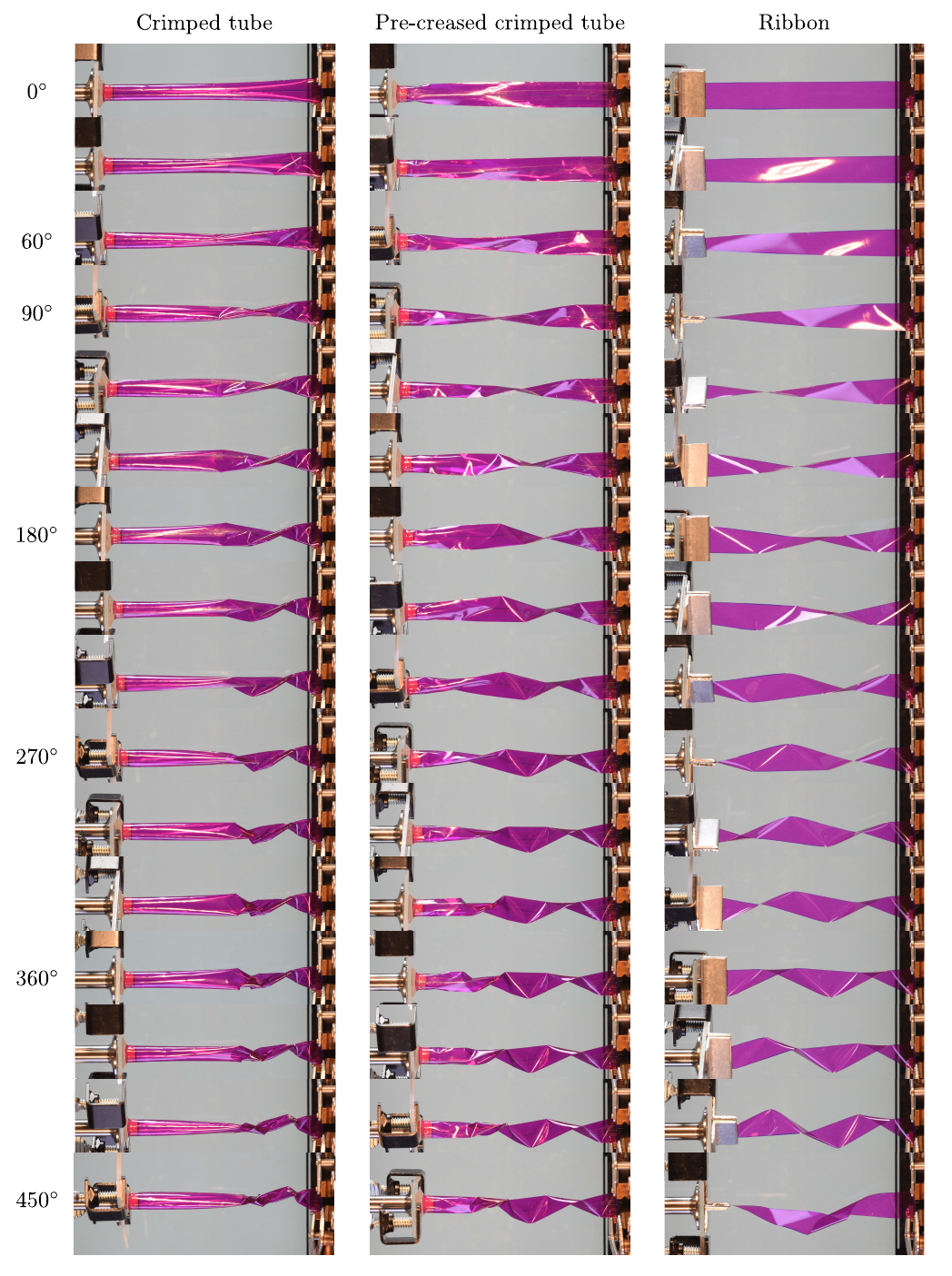}
\caption{\label{fig:sm1}\textbf{Structures during twisting.} Each column is a superset of the images in the corresponding column of Fig.~\ref{fig:360_twist}, taken from the same video of twisting with constant tension. One video frame is shown for every $30^\circ$ of twist, from $0^\circ$ to $450^\circ$.
}
\end{figure*}

\begin{figure}[]
\includegraphics[width=0.9\columnwidth]{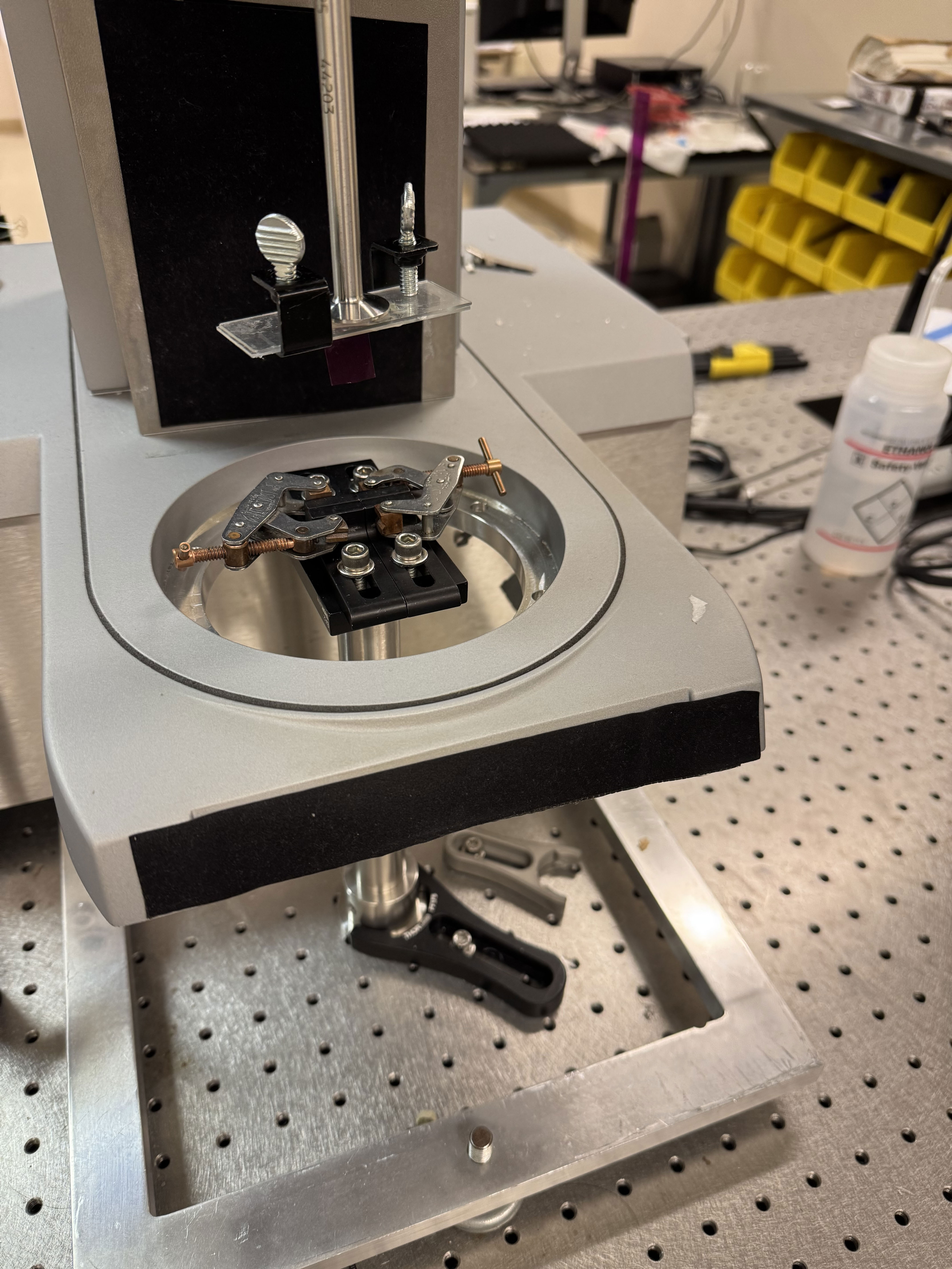}
\caption{\label{fig:setup} \textbf{Method for clamping samples to rheometer.} A glass slide is attached to the upper measuring tool using hot-melt adhesive. This serves as a platform for securing the top of each sample. Two ``L"-shaped clamps hold the bottom crimped end of the samples.  These clamps are secured to the optical table. 
}
\end{figure}

Here we provide additional figures detailing the experiments. 
Figure~\ref{fig:sm1} shows additional images from the experiment in Fig.~\ref{fig:360_twist}. Images are in $30^\circ$ increments from $0^\circ$ to $450^\circ$.
Figure~\ref{fig:setup} shows how we mount our samples to the rheometer.

\begin{figure*}[]
\includegraphics[width=\textwidth]{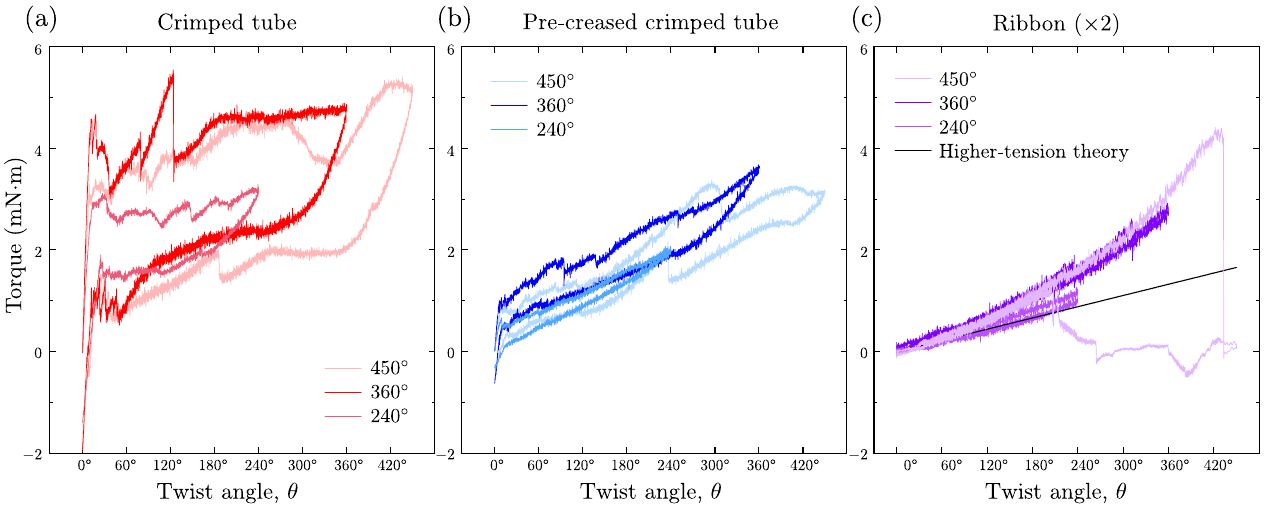}
\caption{\label{fig:3twists} \textbf{Sample variation of torque measurements and dependence of mechanical response on maximum twist angle.} 
Supplementing Fig.~\ref{fig:torque}, here we show torque measurements for multiple maximum twist angles for crimped tubes, pre-creased crimped tubes, and ribbons. 
Thicknesses, dimensions, and applied tensions match those in Figs.~\ref{fig:360_twist},\ref{fig:torque}. 
Samples are twisted from $0^\circ$ to $240^\circ$, $360^\circ$, and $450^\circ$, respectively, and then twisted back to $0^\circ$. 
The torques for the ribbon data and ribbon theory are multiplied by 2, as the ribbon width is half of the circumference of the tubes. 
The large drop at $\theta>420^\circ$ for the ribbon occurs when its morphology suddenly transitions from a creased helicoid to a looped structure \cite{Chopin2013,Demery2018}. 
Black line: Theoretical prediction for the ribbon in the ``far-from-threshold'' regime of Chopin and Filho~\cite{Chopin2019}, multiplied by two. 
The plotted slope $W^2 T/(2L)$ should obtain for tensions $\sim$10 times higher than in our experiments, i.e.\ for $T/(EtW) \gtrsim 2 \times 10^{-3}$. 
Our observed slope is roughly double this higher-tension prediction, and rises with increasing $\theta$, consistent with experiments in that work.
}
\end{figure*}

Figure~\ref{fig:3twists} provides more measurements with different maximum twist angles and compares them with the data from Fig.~\ref{fig:torque}. Whereas sample variation is clearly present in the location and size of the stress drops, the main qualitative trends highlighted in the main text are supported by these additional data -- namely, the magnitude of the stress drops and the hysteresis tend to be largest for the crimped tube, intermediate for the pre-creased crimped tube, and smallest for the ribbon. In twisting ribbons, we observed the transition from a creased helicoid to a looped structure~\cite{Chopin2013,Demery2018}, which corresponds to the large drop in the ribbon data at $\theta > 420^\circ$ in Fig.~\ref{fig:3twists}.

%

\end{document}